\documentclass[aps,draft,eqsecnum,showpacs]{revtex4}
\input epsf
\pacs{05.10.-a, 05.30.-d, 72.15.Rn}

\newcommand{\be}{\begin{equation}}
\newcommand{\ee}{\end{equation}}
\newcommand{\bea}{\begin{eqnarray}}
\newcommand{\eea}{\end{eqnarray}}

\begin{document}

\title{\Large Single Superfield Representation for Mixed Retarded and Advanced
Correlators in Disordered Systems}

\author{Daniel G.\ Barci\footnote{Regular Associate of the ``Abdus Salam international centre for theoretical physics, ICTP'', Trieste, Italy.}}

\affiliation{Departamento de F\'\i sica Te\'orica,
Universidade do Estado do Rio de Janeiro,Rua S\~ao Francisco Xavier 524, 20550-013,  
Rio de Janeiro, RJ, Brazil.}

\author{Luis E.\ Oxman}

\affiliation{Instituto de F\'{\i}sica, Universidade Federal Fluminense,
Campus da Praia Vermelha, Niter\'oi, 24210-340, RJ, Brazil.}

\date{August 9, 2004}

\begin{abstract}
We propose a new single superfield representation for mixed retarded and advanced 
correlators for noninteracting disordered systems. The method is tested in the simpler 
context of Random Matrix theory, by comparing with well known universal behavior for 
level spacing correlations. Our method is general and could be especially interesting
to study localization properties encoded in the mixed correlators of Quantum Hall 
systems.
\end{abstract}

\maketitle

\section{Introduction \label{s1}}

The physics of electronic localization is one of the most interesting issues
in the context of disordered quantum systems \cite{Anderson}. 
The localization/delocalization transition 
has been intensively studied by following the field theoretical approach of Wegner
\cite{Wegner}, 
and the more formal developments by McKane and Stone \cite{Stone}. Universal aspects
of this 
transition and related topics like transport phenomena in disordered media can also
be studied in 
the framework of random matrix theory (for a review see, for instance,
ref. \cite{Mirlin}).

In an integer quantum Hall system, electron localization plays a fundamental role
\cite{Laughlin}, namely, the effect of disorder is that of localizing the
states of energy $E$
away from the Landau energy $E_n=(n+1/2)\hbar\omega_c$. Near this value, the
localization length 
$\xi(E)$ diverges like $\xi(E)\propto (E-E_n)^{-\nu}$. 
Although this behavior is consistent with experiments \cite{exp} and with several 
numerical \cite{numerical} and renormalization group calculations \cite{Moriconi}, 
to the best of our knowledge, there is no simple analytical method to evaluate this
scaling 
\cite{Moore} nor it is clear why, in the thermodynamic limit, just one state for
each Landau band remains extended.

The electronic properties are encoded in the system's correlators; for instance, 
the density of states (localized or not) can be obtained
from the retarded Green's function,
\be
\rho(E)=\frac{1}{\pi} \lim_{\eta\to 0^+} \mbox{\em Im} \langle G(x,x; E-i\eta)\rangle,
\ee
where $\langle ...\rangle$ denotes the ensemble average over the random variable.
On the other hand,  the analytical structure of the density of {\em localized}
states $\rho_\ell(E)$ 
is more complex since it involves the mixed product of retarded and advanced Green's
functions,
\be
\rho_\ell(E)=\frac{1}{\pi}\int dx \lim_{\eta\to 0^+} \eta  
\langle G(x,y; E-i\eta)\;G(x,y; E+i\eta)\rangle.
\label{loc}
\ee
In fact, any quantity encoding some information about localization or transport 
properties is related with this mixing. For this reason, we
concentrate our
efforts in studying the representation of the ensemble average for these important
quantities. 

There are essentially two methods to calculate averages: the replica
\cite{replica} and the supersymmetric method \cite{SUSY}. 
The first one is based on the introduction of a number 
of copies to represent each type of Green's function. In order to compute the
average density of states one set of $n$ field copies is needed.
In this case, a generalized $n$ component Landau-Ginsburg model is 
obtained \cite{Brezin}, 
where the limit $n\to 0$ must be implemented. 
For the density of localized states two sets of copies are used, 
containing $n_+$ and $n_-$ field copies, respectively. The disorder average, after integrating out all massive modes, 
leads to a generalized nonlinear sigma model with symmetry group $O(n_+,n_-)$ 
\cite{Wegner,Stone}, describing the critical properties near the mobility edge.

In the supersymmetric method, the Green's function denominator is exponentiated 
by means of Grassman variables, thus avoiding dangerous continuations in the number
of copies. In order to compute the density of states a superfield is needed, this also applies
to other quantities that can be computed from correlators just involving a retarded prescription.
This method has been used by Br\'ezin, Gross and Itzykson \cite{Gross} to study the density 
of states in  Hall systems. They showed that the reason behind the exact expression previously 
calculated by Wegner \cite{Wegner} is an exact boson-fermion symmetry in the Green's function 
representation, when the projection to the first Landau level is considered.
However, the problem is not so simple when one tries to extend these ideas to understand 
localization. As occurs in the replica method, a doubling is used to represent
the density of localized states, that is, a superfield is introduced 
for each prescription in eq. (\ref{loc}). As a consequence, the boson-fermion symmetry 
is lost, thus precluding the possibility of obtaining exact results 
(see the discussion in ref. \cite{Gross}).

Therefore, it is interesting to look for alternative representations for mixed
correlators, thus opening the possibility to advance the
understanding of disordered systems. 

The aim of this paper is to propose a new representation for mixed correlators 
based on the introduction of essentially just one superfield. 
Our method is general and can be applied to any noninteracting disordered system. 
It is based on the introduction of a single operator, 
quadratic in the system's Hamiltonian, that permits to encode the nontrivial 
information about retarded-advanced mixing by using a single prescription.
Then, the basis of the method is in fact quite general and could be useful outside the
supersymmetric framework.

Our method will be tested in the simpler context of random matrix theory where the supersymmetric method has been introduced by the pioneering work of Efetov \cite{Efetov}.
We will also show how the boson-fermion symmetry can be restored when 
considering mixed correlators in Hall systems. 

In section  \ref{s2}, we outline the general single superfield method for mixed
correlators. In  sections \ref{s3} and \ref{s4} we compute the corresponding  
averages for the Gaussian Unitary Ensemble, while in section \ref{s5} we study 
level spacing correlators and compare with well known universal asymptotic 
behavior. In section \ref{bfs} we briefly discuss quantum Hall (QH) systems. 
Finally,  section \ref{s6} is devoted to present our conclusions and discuss possible 
perspectives.

\section{Single superfield/supervector representation}
\label{s2}

We start by considering the operator, 
\begin{equation}
O({\cal E})=H-{\cal E},
\end{equation}
where $H$ is the system's Hamiltonian. We will use the notation,
\begin{equation}
O_{1+}=O(E_1+i\eta)\makebox[.7in]{and} O_{2-}=O(E_2-i\eta),
\label{Oud}
\end{equation}
where $\eta > 0$ and $E_1$, $E_2$ are two real energies. 
Now, we will define the operator $O_m$, which is quadratic in $H$
and mixes the retarded and advanced prescriptions,
\begin{eqnarray}
O_{m}&=&O_{1+}\, O_{2-}=O_{2-}\, O_{1+}
\label{comm} \\ &=&(H-E)^2-(r/2+i\eta)^2,
\label{mm1}
\end{eqnarray}
where $E=(E_1+E_2)/2$ and $r=E_1-E_2$. As the factors in $O_{m}$ commute, 
it is easy to verify,
\begin{equation}
O_{m}\,[O_{1+}^{-1}-O_{2-}^{-1}]
=O_{2-}-O_{1+}=(r+2i\eta),
\end{equation}
therefore, the inverse of the operator $O_{m}$ is,
\begin{equation}
O_{m}^{-1}=\frac{1}{r+2i\eta}\, [O_{1+}^{-1}-O_{2-}^{-1}]
\label{Oinv}
\end{equation}
(for $\eta \neq 0$, these inverses are well defined). In other
words, the three types of associated Green's functions $G(x,y)=\langle x| O^{-1}|y\rangle$
are related by,
\begin{equation}
G_{m}(x,y;E,r)=\frac{1}{r+2i\eta}\, [G(x,y;E_1 -i\eta)-G(x,y;E_2+i\eta)].
\label{mm}
\end{equation}
Our strategy is based on first using a single superfield to compute 
four-point correlators associated with the $O_{m}$ operator. Then, from Wick's
theorem and eq. (\ref{mm}), we can relate this correlator to the sum of three terms 
involving only retarded, only advanced or mixed products of two-point Green's functions,
respectively. The first two terms are in general simpler to study and well known in
many different disordered systems.
Then, we see that the four-point correlators for $O_{m}$ operators encode important 
information about mixed products. Moreover, as the retarded-retarded and advanced-advanced 
terms also admit a single superfield representation, the ensemble average for the mixed prescriptions could be performed on independent single superfield representations.\\

In particular, in the random matrix case, this procedure will be tested by
calculating finite $N$ correlation functions and comparing with well known large $N$ 
universal results for the Gaussian Unitary Ensemble (GUE), given by the probability
distribution,
\begin{equation}
P(H)dH={\cal N}\prod_{i=1}^N dH_{ii}\prod_{i<j}dH_{ij}\, d\overline{H}_{ij}
\exp (-\frac{1}{2}\gamma N \,tr (H^2)),
\label{proba}
\end{equation}
where $H$ is an $N\times N$ hermitian matrix. 

Taking traces in eq. (\ref{Oinv}) we get,
\begin{equation}
\,tr(O_{m}^{-1})=[\,tr (O_{1+}^{-1})-\,tr (O_{2-}^{-1})]/(r+2i\eta).
\label{tOinv}
\end{equation}
Now we can multiply eq. (\ref{Oinv}) twice, take the trace, multiply
eq. (\ref{tOinv}) twice and then sum together the two expressions to obtain,
\begin{eqnarray}
\lefteqn{\,tr(O_{1+}^{-1})\,tr
(O_{2-}^{-1})+\,tr(O_{1+}^{-1}O_{2-}^{-1})=\nonumber}\\
&&=1/2 [\,tr(O_{1+}^{-1})\,tr(O_{1+}^{-1})+\,tr(O_{1+}^{-1}O_{1+}^{-1})+
\,tr(O_{2-}^{-1})\,tr(O_{2-}^{-1})+\,tr(O_{2-}^{-1}O_{2-}^{-1})]\nonumber \\
&&-(r+2i\eta)^2/2[\,tr(O_{m}^{-1})
\,tr(O_{m}^{-1})+\,tr(O_{m}^{-1}O_{m}^{-1})].
\label{rep}
\end{eqnarray}

When $\eta$ goes to zero, the first member in eq. (\ref{rep}) is a mixed retarded 
and advanced four-point correlator at energies $E_1$ and $E_2$, respectively.

On the other hand, in the second member, we have the sum of three four-point
correlators for the operators $O_{1+}$, $O_{2-}$ and $O_{m}$,
respectively. For each correlator we can introduce a single supervector 
representation of the type,
\begin{eqnarray}
\lefteqn{\,tr(O^{-1})\,tr(O^{-1})+\,tr(O^{-1}O^{-1})=\nonumber}\\
&&(\lambda/2)^2(-1)^{N(N+1)/2}\pi^{-N}\int d\varphi\,
d\overline{\varphi}\,d\psi\,d\overline{\psi}\, 
(\varphi \cdot \overline{\varphi})^2 \exp (-S),
\label{4pc}
\end{eqnarray}
\begin{equation}
S=\frac{\lambda}{2}(\overline{\varphi}O\varphi+\overline{\psi}O\psi),
\label{Sact}
\end{equation}
where $\varphi$ ($\psi$) is a bosonic (fermionic) $N$-component complex
vector. The parameter $\lambda$ must be chosen for the integrals to be
well defined.

Then, the idea is using three independent single supervector representations of the
type given in eq. (\ref{4pc}) to compute the average for the mixed correlator. 
This is in contrast with the usual
supervector doubling to represent a typical term in the first member of eq.
(\ref{rep}),
\begin{eqnarray}
\lefteqn{O_{1+}^{-1}|_{ij} O_{2-}^{-1}|_{kl}=(1/2)^2 \pi^{-2N}\int d\varphi_+\, 
d\overline{\varphi}_+\,d\psi_+\,d\overline{\psi}_+\, 
d\varphi_-\,d\overline{\varphi}_-\,d\psi_-\,d\overline{\psi}_-\,\,
\varphi_{+i}\overline{\varphi}_{+j} \varphi_{-k}\overline{\varphi}_{-l}} \times
\nonumber \\
&&\times \exp \left(-\frac{i}{2}(\overline{\varphi}_+ O_{1+}\varphi_+ +
\overline{\psi}_+ O_{1+}\psi_+)
+ \frac{i}{2}(\overline{\varphi}_- O_{2-}\varphi_-+\overline{\psi}_-
O_{2-}\psi_-)\right).\nonumber \\
\end{eqnarray}

\section{Averages for the four-point correlators of $O_{m}$ operators}
\label{s3}

When averaging over the unitary ensemble in the second member of eq. (\ref{rep}), 
the more involved part of the calculation comes from the $O_{m}$ correlator, as it is
associated with an exponent which depends quadratically on $H$. Then, in this section 
we will concentrate on this part, providing a detailed derivation of the correlator, 
\begin{eqnarray}
\lefteqn{\langle \,tr(O_{m}^{-1})
\,tr(O_{m}^{-1})+\,tr(O_{m}^{-1}O_{m}^{-1})\rangle=\nonumber}\\
&&(\lambda/2)^2 (-1)^{N(N+1)/2}\pi^{-N}\int d\varphi\,
d\overline{\varphi}\,d\psi\,d\overline{\psi}\, 
(\varphi \cdot \overline{\varphi})^2 \langle \exp (-S_{m})\rangle,
\label{pri}
\end{eqnarray}
\begin{equation}
S_{m}=\frac{\lambda}{2}\overline{\varphi}\,O_{m}\,\varphi+
\frac{\lambda}{2}\overline{\psi}\,O_{m}\,\psi.
\end{equation}
Using eq. (\ref{mm1}), 
and considering the matrix $K=\varphi \otimes \overline{\varphi} -\psi \otimes
\overline{\psi}$, we can also write,
\[
\langle \exp (-S_{m})\rangle={\cal N}
\exp
\left(\frac{\lambda}{2}[(r/2+i\eta)^2-E^2]\,
tr K\right)
\int dH\, \exp(-\,tr HSH+\lambda E \,tr HK),
\]
where $S=\frac{1}{2}\gamma N I +\frac{\lambda}{2}K $. Completing squares, we have,
\begin{equation}
\langle \exp (-S_{m})\rangle={\cal N}
\exp
\left(\frac{\lambda}{2}[(r/2+i\eta)^2-E^2]
\,tr K+\frac{\lambda^2}{4}E^2\,tr KS^{-1}K\right)
(2/\gamma N)^{\frac{N^2}{2}} I(\tilde{S}),
\label{si}
\end{equation}
where we have defined,
\begin{equation}
I(\tilde{S})=\int dH\, \exp(-\,tr H\tilde{S}H)
\makebox[.7in]{,}
\tilde{S}= I +\frac{\lambda}{\gamma N}K .
\label{til}
\end{equation}
In order to evaluate $I(\tilde{S})$, we will consider the general case
where $\tilde{S}=I + g(K)$, substituting $g(x)=\frac{\lambda}{\gamma N} x $ at the end 
of the calculation. 

As the measure $dH$ is invariant under unitary transformations 
$H\rightarrow UHU^{\dagger}$, $U\in U(N)$, we have, $I(U\tilde{S}U^{\dagger})=I(\tilde{S})$;
and taking the matrix $U$ that
diagonalizes $\tilde{S}$: $U\tilde{S}U^{\dagger}=D$, $\tilde{S}=U^{\dagger}DU $ we can
write,
\begin{eqnarray}
&&I(\tilde{S})=I(D)=\int dH \exp \left( -\sum_i d_i H_{ii}^2 - \sum_{i<j}(d_i+d_j)
H_{ij}\overline{H}_{ij}\right)\nonumber \\
&&=\pi^{\frac{N^2}{2}}\prod_i d_i^{-1/2}\prod_{i<j}(d_i+d_j)^{-1}=\pi^{\frac{N^2}{2}}2^{\frac{N}{2}}\exp\left(-1/2\sum_{i,j} \ln
(d_i+d_j)\right),\nonumber
\end{eqnarray}
where $d_i=1+g_i$ are the diagonal elements of $D$, written in terms of the eigenvalues 
$g_i$ of the matrix $g(K)$, so we obtain,
\begin{equation}
I(\tilde{S})=
(\pi/2)^{\frac{N^2}{2}}2^{\frac{N}{2}}\exp \left(-1/2\sum_{i,j} \zeta \left(\frac{g_i+g_j}{2}\right)\right)
\makebox[.5in]{,}
\zeta(x)=\ln (1+x).
\label{sim}
\end{equation}
Now, if we expand $\zeta (x)=\sum_{n=1} a_n x^n$, the exponent in $I(\tilde{S})$ can be written in terms 
of $U(N)$ invariants,
\begin{eqnarray}
\sum_{i,j} \zeta \left(\frac{g_i+g_j}{2}\right)&=&\sum_{i,j}\sum_n \frac{a_n}{2^n} \sum_l 
\left( \begin{array}{l} n\\ l\\ \end{array}\right) g_i^l g_j^{n-l}\nonumber \\
&=& \sum_n \frac{a_n}{2^n} \sum_l \left( \begin{array}{l} n\\ l\\ \end{array}\right) tr (g^l) tr(g^{n-l}).
\label{zeta}
\end{eqnarray}
In the appendix, we show that the trace of a given function
$f(K)$, with $K=\varphi \otimes \overline{\varphi} -\psi \otimes \overline{\psi}$, is
given by,
\begin{equation}
\,tr f(K)=Nf_0+f(\alpha)-f(\beta)+\frac{f'(\alpha)-f'(\beta)}{\alpha-\beta}
\mu \overline{\mu},
\label{formu}
\end{equation}
\[
\alpha=\varphi \cdot \overline{\varphi}~~~,~~~
\beta=\psi \cdot \overline{\psi}~~~,~~~
\mu=\varphi \cdot \overline{\psi}.
\]
Using this formula in eq. (\ref{zeta}), we can resume the series in the index $n$,
\begin{eqnarray}
\lefteqn{\sum_{i,j} \zeta (\frac{g_i+g_j}{2})=}\nonumber \\
&&\zeta(g(\alpha)) +2N \zeta \left(g(\alpha)/2\right)+\zeta(g(\beta))- 2N \zeta (g(\beta)/2)-2 \zeta(g(\alpha)/2+g(\beta)/2) +
\nonumber \\
&&+ \frac{\mu \overline{\mu}}{\alpha -\beta}\, \frac{d~}{d\alpha}\left[\zeta(g(\alpha))+ 2N  \zeta(g(\alpha)/2) -2 \zeta(g(\alpha)/2+g(\beta)/2) \right]+\nonumber \\
&&+ \frac{\mu \overline{\mu}}{\alpha -\beta}\, \frac{d~}{d\beta}\left[\zeta(g(\beta))- 2N  \zeta(g(\beta)/2) -2 \zeta(g(\alpha)/2+g(\beta)/2) \right],
\label{zinc}
\end{eqnarray}
where we have used $g(0)=0$ and $\zeta(0)=0$.

Now, from eqs. (\ref{sim}), (\ref{zinc}), and also considering the formula (\ref{formu})
in the first factor of eq. (\ref{si}), we finally obtain,
\begin{equation}
\langle \exp (-S_{m})\rangle=G_1(\alpha,\beta)+G_2(\alpha,\beta)\mu
\overline{\mu},
\label{proms}
\end{equation}
where,
\begin{eqnarray}
G_1&=&\frac{1}{2}
\exp \left( \frac{\lambda}{2}\left[
(r/2+i\eta)^2-E^2\right](\alpha-\beta)+\frac{\lambda^2 E^2}{2}
\left[ \frac{\alpha^2}{\gamma N+\lambda \alpha}-\frac{\beta^2}{\gamma
N+\lambda \beta}\right]\right)\times \nonumber \\
&&\times \frac{(2+\lambda \beta/\gamma N)^N}{(2+\lambda \alpha/\gamma
N)^N}\frac{(2+\lambda (\alpha+\beta)/\gamma N)}{(1+\lambda \alpha/\gamma
N)^{1/2}(1+\lambda \beta/\gamma N)^{1/2}},
\label{G1}
\end{eqnarray}
and
\begin{eqnarray}
G_2&=&G_1  \left\{\frac{\lambda^2 E^2}{2\gamma
N}\frac{(2+\lambda (\alpha 
+\beta)/\gamma N)}{(1+\lambda \alpha/\gamma
N)^2(1+\lambda 
\beta/\gamma N)^2} +
\frac{\lambda^2}{\gamma^2 N
(2+\lambda \alpha/\gamma N)(2+\lambda \beta/\gamma N)}+\right.
\nonumber \\
&&\left. +\frac{\lambda^3 (\beta-\alpha)}{2\gamma^3 N^3 
(1+\lambda \alpha/\gamma N)(1+\lambda \beta/\gamma N)(2+\lambda
(\alpha+\beta)/\gamma N)}\right\}.
\label{G2}
\end{eqnarray}
Note that this expression is correctly normalized as for $\varphi=0$,
$\psi=0$ we have, $S_{m}=0$, $\alpha=0$, $\beta=0$ and $\mu=0$. Then, in this case,  
we see from eq. (\ref{G1}) that the first and second members in eq. (\ref{proms})
are equal to one.

Now, we can substitute eq. (\ref{proms}) in the correlator (\ref{pri}) to obtain,
\begin{eqnarray}
\lefteqn{\langle \,tr(O_{m}^{-1})
\,tr(O_{m}^{-1})+\,tr(O_{m}^{-1}O_{m}^{-1})\rangle=\nonumber}\\
&&(\lambda/2)^2(-1)^{N(N+1)/2}\pi^{-N}\int d\varphi\,
d\overline{\varphi}\,d\psi\,d\overline{\psi}\, 
\alpha^2 [G_1(\alpha,\beta)+G_2(\alpha,\beta)\mu
\overline{\mu}].\nonumber \\
\end{eqnarray}
We will still transform this expression to a function of 
$\alpha$ and $\beta$ only. In order to do so, we
recall that the fermionic integral picks-up the term of the integrand
having the form $D\overline{\psi}_1\psi_1\overline{\psi}_2 \psi_2\dots
\overline{\psi}_N\psi_N$. Now, in $\beta=\psi \cdot \overline{\psi}$,
the $\psi_i$'s and $\overline{\psi}_j$'s come in pairs
$\psi_i\overline{\psi}_i$; then, the crossed terms ($i\neq j$) in
\[
\mu \overline{\mu}=\varphi \cdot \overline{\psi}\,\,
\overline{\varphi} \cdot \psi 
=\sum_i\varphi_i \overline{\varphi}_i \overline{\psi}_i \psi_i 
+\sum_{i\neq j}\varphi_i \overline{\varphi}_j \overline{\psi}_i \psi_j,
\]
cannot contribute to the fermionic integral, that is,
\begin{eqnarray}
&&\int d\psi\,d\overline{\psi}\, G_2(\alpha,\beta)\mu \overline{\mu}\int d\psi\,d\overline{\psi}\, G_2(\alpha,\beta)
\sum_i\varphi_i \overline{\varphi}_i \overline{\psi}_i \psi_i \nonumber \\
&&=\frac{1}{N}\sum_i \varphi_i \overline{\varphi}_i
\int d\psi\,d\overline{\psi}\, G_2(\alpha,\beta) \overline{\psi}
\cdot \psi =\frac{1}{N}\varphi \cdot \overline{\varphi}
\int d\psi\,d\overline{\psi}\, G_2(\alpha,\beta) \overline{\psi}
\cdot \psi,\nonumber \\
\label{intgm}
\end{eqnarray}
where we have also used the symmetry among the $\psi_i$'s. Putting this information
together, we arrive at,
\begin{eqnarray}
\lefteqn{\langle \,tr(O_{m}^{-1})
\,tr(O_{m}^{-1})+\,tr(O_{m}^{-1}O_{m}^{-1})\rangle=\nonumber}\\
&&(\lambda/2)^2(-1)^{N(N+1)/2}\pi^{-N}\int d\varphi\,
d\overline{\varphi}\,d\psi\,d\overline{\psi}\, 
\alpha^2 [G_1(\alpha,\beta)-\frac{\alpha \beta}{N} G_2(\alpha,\beta)].\nonumber \\
\end{eqnarray}
Now, the integrals over $\varphi$, $\overline{\varphi}$ can
be easily expressed in terms of a single variable  integral over $\alpha$. Using the 
solid angle in $M$ dimensions, $\Omega_{M}=2\pi^{M/2}/\Gamma(M/2)$,  
and the fermionic integral,
\[
\int d\psi
d\overline{\psi}\,F(\beta)=(-1)^{N(N+1)/2}\left.\frac{d^N}{d\beta^N}F 
\right|_{\beta=0},
\]
we finally obtain,
\begin{eqnarray}
\lefteqn{\langle \,tr(O_{m}^{-1})
\,tr(O_{m}^{-1})+\,tr(O_{m}^{-1}O_{m}^{-1})\rangle=\nonumber}\\
&& \frac{(\lambda/2)^2}{(N-1)!}\int_0^{\infty} d\alpha\,
\alpha^{N+1}\frac{d^N}{d\beta^N}  \left[ G_1(\alpha,\beta)-\frac{\alpha
\beta}{N}
G_2(\alpha,\beta)\right]_{\beta=0}.
\label{TE}
\end{eqnarray}

\section{Analytical determination for the correlators of $O_{m}$ operators}
\label{s4}

Here, we will study the analytical properties of the four-point $O_m$ correlators.
Recalling that $\eta$ is a small parameter 
(we will take the limit $\eta \rightarrow 0$), we can drop $\eta^2$ 
in the exponent $\frac{\lambda}{2}(r/2+i\eta)^2 \alpha$, in eqs. (\ref{G1}) and (\ref{G2}). 
We will suppose $r=E_1-E_2>0$; 
then, for the integral representation (\ref{TE}) be well defined, 
we must take $\lambda=i$. Precisely, this choice produces
an associated factor in the bosonic integral,
\begin{equation}
\exp i\omega\alpha \,\, \exp(-\eta r\alpha/2)
\makebox[.5in]{,}
\omega=\frac{r^2}{8},
\label{ome}
\end{equation}
which leads to a convergent integral. 

We underline here that, as usual, the factor $N$ in the exponent of the probability measure in eq. (\ref{proba}) assures that the energy band attain a finite width when large $N$ matrices are considered. Therefore, the natural scale for the level spacing is $1/N$. For this reason, we will define the variables,
\begin{equation}
r=\frac{s}{N} \makebox[.5in]{,} \gamma N \omega=\frac{\chi}{N}
\makebox[.5in]{,}
\chi=\gamma \frac{s^2}{8}.
\label{ochi}
\end{equation}
Using eqs. (\ref{G1}), (\ref{G2}) and (\ref{TE}), working around  $E=(E_1+E_2)/2=0$,
and considering the change of variables $\alpha \to \alpha/\omega$, $\beta \to -i\beta/\omega$,
the small $\eta$ ($\eta << r$) contribution of the last term in eq. (\ref{rep}) to the mixed 
retarded and advanced correlator results,
\begin{eqnarray}
\lefteqn{C(\chi)=-\frac{r^2}{2} \langle \,tr(O_{m}^{-1})
\,tr(O_{m}^{-1})+\,tr(O_{m}^{-1}O_{m}^{-1})\rangle =\nonumber} \\
&&=\gamma N^2 i^{N} \int_0^{\infty} d\alpha\,
\alpha^{N+1}\exp(i\alpha)\left. \frac{1}{N!}\frac{d^N}{d\beta^N}\right|_{\beta=0}
 \exp(-\beta)\times \nonumber \\
&&\times\left[
\frac{(\frac{\chi}{N}+ b)^{N-1}}{(\frac{\chi}{N} +a)^{N+1}}
\frac{(a+b)^2}{(ab)^{1/2}} -
\frac{1}{4\chi}\frac{(\frac{\chi}{N}+b)^{N}}{(\frac{\chi}{N} +a)^{N}}
\frac{(a-\frac{\chi}{N})(b-\frac{\chi}{N})(b-a)}{(ab)^{3/2}}
\right],\nonumber \\
\label{cderre}
\end{eqnarray}
where $a=\frac{\chi}{N}+i\alpha$, $b=\frac{\chi}{N}+ \beta$.

We see that this expression is a sum of terms containing products of the form
$I^{(n)}_{\nu} Q^{(n)}_{\mu}$, with $n=0,1$ and $\mu$, $\nu$ semi-integers, where,
\begin{equation}
I^{(n)}_{\nu}(\chi)=\gamma i^{N} \int_0^{\infty} d\alpha\, 
\frac{\alpha^{N+1}(\frac{\chi}{N}+i\alpha)^\nu\exp(i\alpha)}{(\frac{2\chi}{N} + i\alpha)^{N+1-n}} ,
\label{fact}
\end{equation}
and the $Q$'s are polynomials,
\begin{equation}
Q^{(n)}_\mu(\chi)=\left. \frac{1}{N!}\frac{d^N}{d\beta^N}\right|_{\beta=0}(\frac{\chi}{N}+ b)^{N-1+n}b^{\mu}\exp(-\beta).
\label{polyQ}
\end{equation}
Now, the integrand in eq. (\ref{fact}) presents singularities on the upper
complex $\alpha$-plane, on the imaginary axis. There is an isolated
singularity when $\alpha=2i \chi/N$, and there
is a continuum of singularities for $\alpha=iy$, $y>\chi/N$. 
As no singularities are found on the first
quadrant, and the $\alpha$ integral on the arc at infinity goes to
zero, we can rotate the integration path until the positive imaginary axis is approached 
anti-clockwise. Recalling that, from eq. (\ref{ome}), $\omega$ contains a small positive imaginary part that regularizes the integral, according to eq. (\ref{ochi}) we have also to consider, $\chi \rightarrow \chi +i\epsilon$. Then, we can use the equivalent representation,
\begin{equation}
I^{(n)}_\nu(\chi)=\gamma (-1)^{N+1} \int_0^{\infty} dy\, 
\frac{y^{N+1}(\frac{\chi}{N}-y +i\epsilon )^\nu\exp(-y)}{(\frac{2\chi}{N}-y + i\epsilon)^{N+1-n}}
\end{equation}
We can split this integral into two parts, 
\begin{equation}
I^{(n)}_\nu = J^{(n)}_\nu + K^{(n)}_\nu,
\label{split}
\end{equation}
where
the integral in the $J$'s runs from $0$ to $\chi/ N$, while the integral 
in the $K$'s runs from $\chi/N$ to $\infty$. A straightforward calculation leads to,
\[
J^{(n)}_\nu (\chi)=
\gamma (-1)^{N+1}(\chi/N)^{n+\nu +1} \int_0^{1} dv\, 
\frac{v^{N+1}(1-v)^\nu \exp(-\frac{\chi}{N}v)}{(2-v)^{N+1-n}}.
\]
On the other hand, we can write,
\[
K^{(n)}_\nu(\chi)=\gamma (-1)^{N+1}e^{i\pi \nu} \int_{\frac{\chi}{N}}^{\infty} dy\,\, 
y^{N+1}(y-\frac{\chi}{N})^\nu\exp(-y) \frac{1}{(N-n)!}\frac{d^{N-n}}{dy^{N-n}}
\left( \frac{1}{\frac{2\chi}{N}-y + i\epsilon}\right).
\]
Note that the factor $e^{i\pi \nu}$ is pure imaginary, as the $\nu$'s are semi-integer. Then, using   
$(2\chi/N-y + i\epsilon)^{-1}= P (2\chi/N-y)^{-1} -i\pi \delta(2\chi/N-y)$, we see that the real part 
of the $K$'s can be easily computed, as this is associated with the $\delta$ term,
\begin{equation}
\Re\, K^{(n)}_\nu(\chi)=e^{i\pi (\nu-n+\frac{1}{2})}\gamma \pi 
\frac{1}{(N-n)!}\left. \frac{d^{N-n}}{dy^{N-n}}\right|_{y=\frac{2\chi}{N}}
y^{N+1}(y-\frac{\chi}{N})^\nu\exp(-y), 
\end{equation}
and changing variables, $y=2\chi/N +\beta$,
\begin{equation}
\Re\, K^{(n)}_\nu(\chi)=e^{i\pi (\nu-n+\frac{1}{2})} \, P^{(n)}_\nu(\chi),
\label{guen}
\end{equation}
where we have introduced the functions, 
\begin{equation}
P^{(n)}_\nu(\chi)=\gamma \pi \exp{(-2\chi/N)}\, \frac{1}{(N-n)!}\left. \frac{d^{N-n}}{d\beta^{N-n}}\right|_{\beta=0}
(\frac{\chi}{N}+ b)^{N+1} b^\nu \exp(-\beta), 
\label{qP}
\end{equation}
and $b$ has been defined after eq. (\ref{cderre}).

\section{Correlations for level spacing}
\label{s5}

In this section, in order to test our single superfield method, we will first obtain an approximate relationship 
between the $O_m$-correlators and the DOS-DOS correlators, which becomes exact in the $N\to \infty$
limit. Then, we will compare the finite $N$  $O_m$-correlators, as $N$ becomes large, with the well known universal results for DOS-DOS correlations. 

The density of states operator, at energy $E_1$, is given by,
\begin{equation}
\hat{\rho}(E_1)=\frac{1}{N} \sum_i \delta(E_1-E_i)= \frac{1}{2\pi iN} tr (O_{1+}^{-1}-O_{1-}^{-1}),  
\end{equation}
(cf. eq. (\ref{Oud})). Then, the DOS-DOS correlator can be written as,
\begin{eqnarray}
\lefteqn{\rho(E_1,E_2)=\langle \hat{\rho}(E_1)\hat{\rho}(E_2)\rangle - \langle \hat{\rho}(E_1)\rangle \langle \hat{\rho}(E_2)\rangle =}\nonumber \\
&& \frac{1}{2\pi^2 N^2} \Re \left( \langle tr (O_{1+}^{-1})\, tr (O_{2-}^{-1})\rangle - 
\langle tr (O_{1+}^{-1})\rangle \langle tr (O_{2-}^{-1})\rangle\right)\nonumber \\
&& - \frac{1}{2\pi^2 N^2} \Re \left( \langle tr (O_{1+}^{-1})\, tr (O_{2+}^{-1})\rangle - 
\langle tr O_{1+}^{-1}\rangle \langle tr O_{2+}^{-1}\rangle\right).
\label{dd}
\end{eqnarray}
As we are interested in analyzing the new single superfield representation for mixed correlators, 
we will simplify the standard part of the calculation as much as possible.    
In particular, in the $N\to \infty$ limit, the following factorization of retarded-retarded correlators
is verified (see for example refs. \cite{Mirlin}, \cite{Fyod}),
\begin{equation}
\langle tr (O_{1+}^{-1})\, tr (O_{2+}^{-1})\rangle = 
\langle tr O_{1+}^{-1}\rangle \langle tr O_{2+}^{-1}\rangle.
\label{factoriza}
\end{equation}
In this manner, up to irrelevant terms that scale to zero for large values of $N$, we can consider the 
approximation,
\begin{eqnarray}
\lefteqn{\rho(E_1,E_2)\approx}\nonumber \\
&& \frac{1}{2\pi^2 N^2} \Re \left( -r^2/2\langle \,tr(O_{m}^{-1})
\,tr(O_{m}^{-1})+\,tr(O_{m}^{-1}O_{m}^{-1})\rangle \right) + \frac{R}{2\pi^2 N^2},
\label{roud}
\end{eqnarray}
\begin{eqnarray}
\frac{R}{N^2} &\approx& \frac{1}{N^2}\Re \left(\frac{1}{2} \langle \,tr(O_{1+}^{-1}) \rangle^2
+\frac{1}{2}\langle \,tr(O_{2-}^{-1})\rangle^2 -\langle
\,tr(O_{1+}^{-1})\rangle \langle \,tr(O_{2-}^{-1})\rangle \right)+
\nonumber \\
&& + \frac{1}{N^2}\Re \left( \frac{1}{2} \langle \,tr(O_{1+}^{-1}O_{1+}^{-1})\rangle+\frac{1}{2}\langle \,tr(O_{2-}^{-1}O_{2-}^{-1})\rangle
 -\langle \,tr(O_{1+}^{-1}O_{2-}^{-1})\rangle\right),\nonumber \\
\label{Rinter}
\end{eqnarray}
where we have used eqs. (\ref{rep}), (\ref{factoriza}), and $r\neq 0$, $\eta \to 0$.

In addition, we have,
\begin{eqnarray}
\langle \,tr(O_{1+}^{-1}O_{1+}^{-1})\rangle & =& \langle\sum_i (E_i-E_1-i\eta)^{-2}\rangle=
\frac{d~}{dE_1}\langle \,tr(O_{1+}^{-1})\rangle, \nonumber \\
 \langle \,tr(O_{2-}^{-1}O_{2-}^{-1})\rangle &=& \frac{d~}{dE_2}\langle \,tr(O_{2-}^{-1})\rangle,
\label{prim}
\end{eqnarray}
while, using eq. (\ref{tOinv}) for $r\neq 0$, the last term in eq. (\ref{Rinter}) can be separated in only retarded 
and only advanced parts,
\begin{equation}
\langle \,tr(O_{1+}^{-1}O_{2-}^{-1})\rangle = \frac{1}{r}\langle \,tr (O_{1+}^{-1})\rangle-\frac{1}{r} 
\langle\,tr (O_{2-}^{-1})\rangle.
\label{seg}
\end{equation}
Note that, using (\ref{dd}), (\ref{rep}), (\ref{prim}) and (\ref{seg}), the density-density correlator can be generally written in terms of four-point $O_m$ correlators, up to
only retarded and only advanced terms which can be studied by following standard procedures.

In particular, in the Random Matrix case, we quote the well known large $N$ results,
\[
\frac{1}{N}\, tr(O_{1+}^{-1}) = -\frac{\gamma}{2}E_1 + i\pi \rho (E_1)
\makebox[.5in]{,}
\frac{1}{N}\, tr(O_{2-}^{-1}) = -\frac{\gamma}{2}E_2 - i\pi \rho (E_2),
\]
\begin{equation}
\rho(E)=\frac{\gamma}{2\pi}\sqrt{\frac{4}{\gamma}-E^2}
\makebox[.5in]{,} E^2<\frac{4}{\gamma}. 
\label{dos}
\end{equation}
Then, using eqs. (\ref{prim}) and (\ref{seg}), we see that the three terms in the second line of eq. (\ref{Rinter}), when multiplied by the $1/N^2$ factor, are irrelevant. In fact, the second line tends to zero faster than $1/N$, as from 
eqs. (\ref{prim}) and (\ref{seg}),
\[
\frac{1}{N} \Re \langle \,tr(O_{1+}^{-1}O_{1+}^{-1})\rangle \approx -\frac{\gamma}{2}
\makebox[.5in]{,}
\frac{1}{N} \Re \langle \,tr(O_{2-}^{-1}O_{2-}^{-1})\rangle \approx -\frac{\gamma}{2},
\]
\[
\frac{1}{N} \Re \langle \,tr(O_{1+}^{-1}O_{2-}^{-1})\rangle \approx -\frac{\gamma}{2r}(E_1-E_2)=-\frac{\gamma}{2},
\]
thus, implying that a $1/N$ factor multiplying the sum in the parenthesis of this line would already give a vanishing
large $N$ result.

With regard to the first line in eq. (\ref{Rinter}), let us recall that, as explained after eq. (\ref{ome}),
the appropriate scaling to study level spacing is $r=s/N$, with $s$ fixed. This is the natural variable we have used in the mixed correlator in eq. (\ref{cderre}) ($\chi=\gamma s^2/8$), which we are studying at $E_1+E_2=0$, that is, $E_1=
s/(2N)$, $E_2=-s/(2N)$. As a consequence, in the first line of eq. (\ref{Rinter}), we can use,
$(1/N)\langle \,tr(O_{1+}^{-1})\rangle \approx i\pi \rho (0)$ and 
$(1/N)\langle \,tr(O_{2-}^{-1})\rangle \approx -i\pi \rho (0)$, that is,
\begin{eqnarray}
\frac{R}{2\pi^2 N^2}&\approx & \frac{1}{2\pi^2}\left( \frac{1}{2} (i\pi \rho(0))^2+ \frac{1}{2} (-i\pi \rho(0))^2
-(i\pi \rho(0))(-i\pi \rho(0))\right)\nonumber \\
&\approx& -(\rho(0))^2.
\end{eqnarray}
Then, using this information in eq. (\ref{roud}), we find,
\begin{eqnarray}
1+ \frac{\rho(E_1,E_2)}{\rho(E_1)\rho(E_2)} &\approx & 1+ \frac{\rho(E_1,E_2)}{(\rho(0))^2}\nonumber \\
&\approx & \frac{1}{2\pi^2 (\rho(0))^2 N^2} \Re \left( -r^2/2\langle \,tr(O_{m}^{-1})
\,tr(O_{m}^{-1})+\,tr(O_{m}^{-1}O_{m}^{-1})\rangle \right),\nonumber \\
&\approx & \frac{1}{2\gamma N^2} \Re \, (C(\chi)),
\label{Omm}
\end{eqnarray}
where we have used $\rho(0)=\sqrt{\gamma}/\pi$ (cf. eq. (\ref{dos})) and eq. (\ref{cderre}). 

In this manner, we see that the large $N$ four-point correlator of $O_m$ operators in eq. (\ref{Omm}) gives the usual $\langle \hat{\rho}(E_1)\hat{\rho}(E_2)\rangle/(\rho(E_1)\rho(E_2))$ correlator. 

In order to test our single superfield representation for mixed correlators, we can compare the large $N$ behavior for the last line in eq. (\ref{Omm}) and the well known ($N\to \infty$) universal behavior for the left member in that equation,
\begin{equation} 
1+ \frac{\rho(E_1,E_2)}{\rho(E_1)\rho(E_2)} \approx 1-\left(  \frac{\sin \sqrt{\gamma}\, s}{\sqrt{\gamma}\, s}  \right)^2
\makebox[.5in]{,}
\label{ulaw}
\end{equation}
(note that $\sqrt{\gamma}\, s=\pi r/\Delta$, where $\Delta=1/(\rho(0)N)$ is the mean level spacing).

As we have seen in section \ref{s4}, $C(\chi)$ is a sum of terms containing products of 
the form $I^{(n)}_{\nu} Q^{(n)}_{\mu}$, with real coefficients. The polynomials $Q^{(n)}_{\mu}$
are real (cf. eq. (\ref{polyQ})), so that $\Re\, C(\chi)$ depends on the real part of the $I$'s, which have been splitted
into two terms in eq. (\ref{split}), by using the $J^{(n)}_{\nu}$ and $K^{(n)}_{\nu}$ functions.
It can be seen that the $J$'s contribution to $\Re\, C(\chi)$ is suppressed for large $N$. 
On the other hand, the real part of the $K$'s has been evaluated in closed form in eq. (\ref{guen}). Putting all this information together, we arrive at,
\begin{eqnarray}
&&\frac{1}{2\gamma N^2} \Re \, (C(\chi))\approx \nonumber \\
&&\approx \frac{1}{2\gamma} \left( P^{(0)}_{3/2}\, Q^{(0)}_{-1/2}-2 P^{(0)}_{1/2}\, Q^{(0)}_{1/2}+
P^{(0)}_{-1/2}\, Q^{(0)}_{3/2}+ \frac{1}{4\chi}P^{(1)}_{-1/2}\, Q^{(1)}_{1/2}+\right. \nonumber \\
&& \left. + \frac{1}{4\chi}P^{(1)}_{1/2}\, Q^{(1)}_{-1/2}+ \frac{1}{4N}P^{(1)}_{-3/2}\, Q^{(1)}_{1/2}-
\frac{1}{4N}P^{(1)}_{1/2}\, Q^{(1)}_{-3/2}-\frac{\chi}{4N^2}P^{(1)}_{-3/2}\, Q^{(1)}_{-1/2}-
\frac{\chi}{4N^2} P^{(1)}_{-1/2}\, Q^{(1)}_{-3/2}\right).\nonumber \\
\label{theq}
\end{eqnarray}

\begin{figure}
\vspace{1cm}
\noindent
\hspace{-2.0 cm}
\epsfxsize=6 cm
\epsfysize=6 cm
\epsfbox{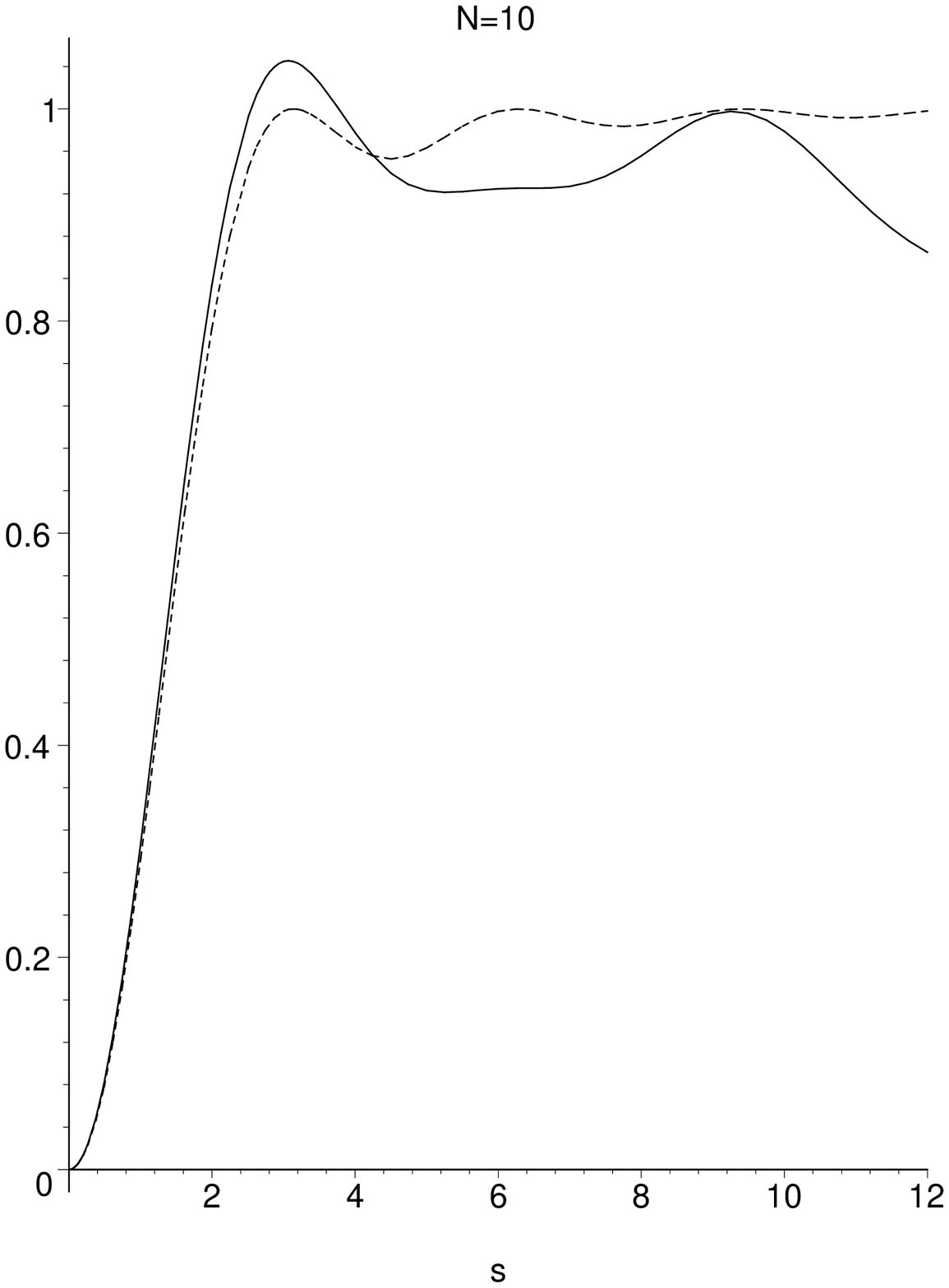}
\vspace{1 cm}
\hspace{3.0 cm}
\epsfxsize=6 cm
\epsfysize=6 cm
\epsfbox{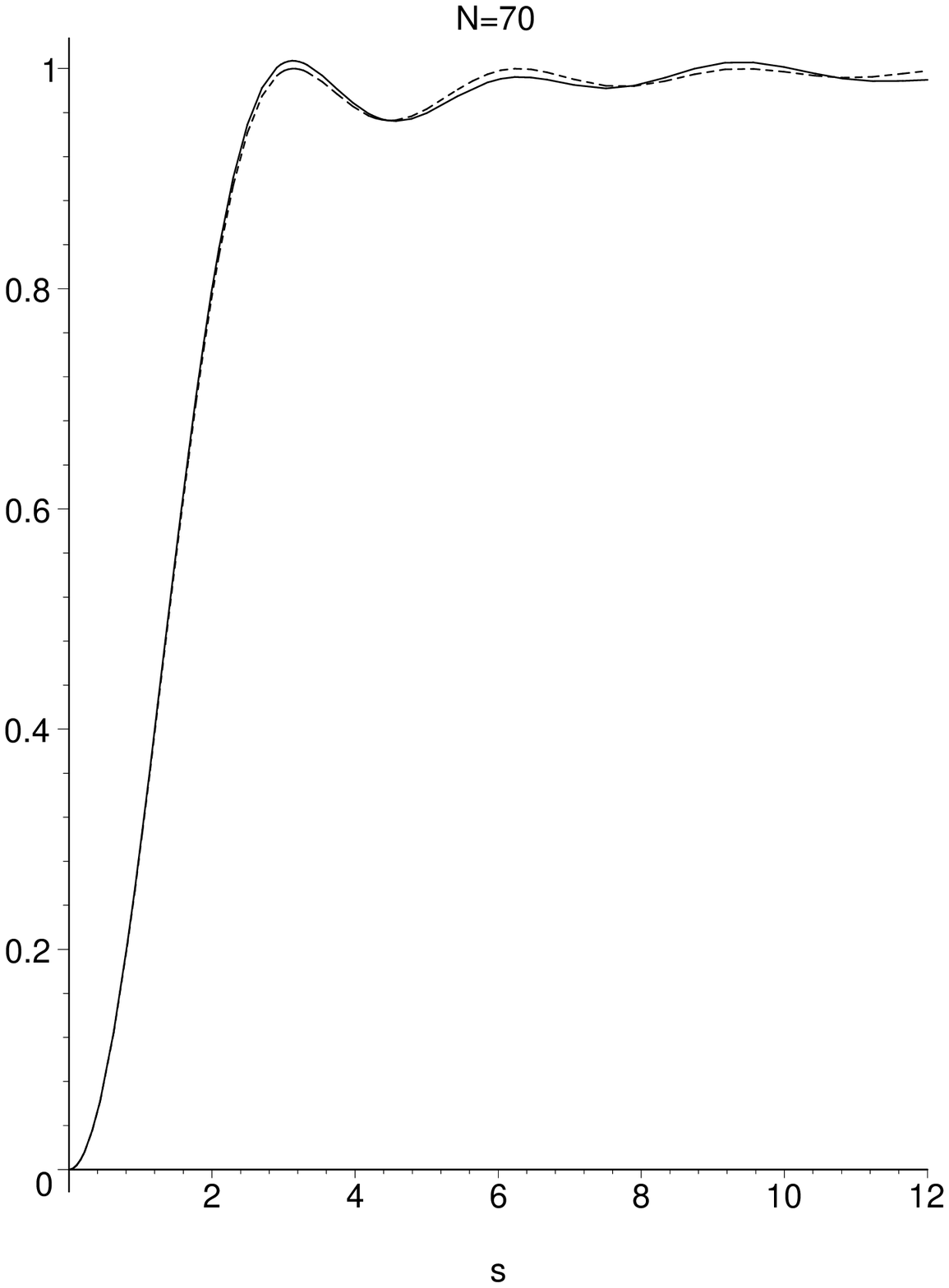}

\caption{In these two figures the bold line represents the correlator (\ref{theq}), for $N=10$ and $N=70$ respectively. The dashed line corresponds to the 
$1-(\sin^2 s)/s^2$ law in eq. (\ref{ulaw}).  We have considered $\gamma=1$ in the two figures}
\label{fig1}
\end{figure}

In figure \ref{fig1}, we display, in the range $s\in [0,12]$, the $N\to \infty$ universal law in eq. (\ref{ulaw}) and our $N=10$ and $N=70$ correlator $(2\gamma N^2)^{-1}\, \Re \, (C(\chi))$ 
in eq. (\ref{theq}). We see that the correlators approach the correct asymptotic form as $N$ increases for all values of the scaled variable $s$.  
For small values of $s$ the asymptotic law is well approximated even with small values of $N$. As $N$ increases, larger values of $s$ progressively fit the $N\to\infty$ behavior.  

\begin{figure}
\vspace{1cm}
\noindent
\hspace{-2.0 cm}
\epsfxsize=6 cm
\epsfysize=6 cm
\epsfbox{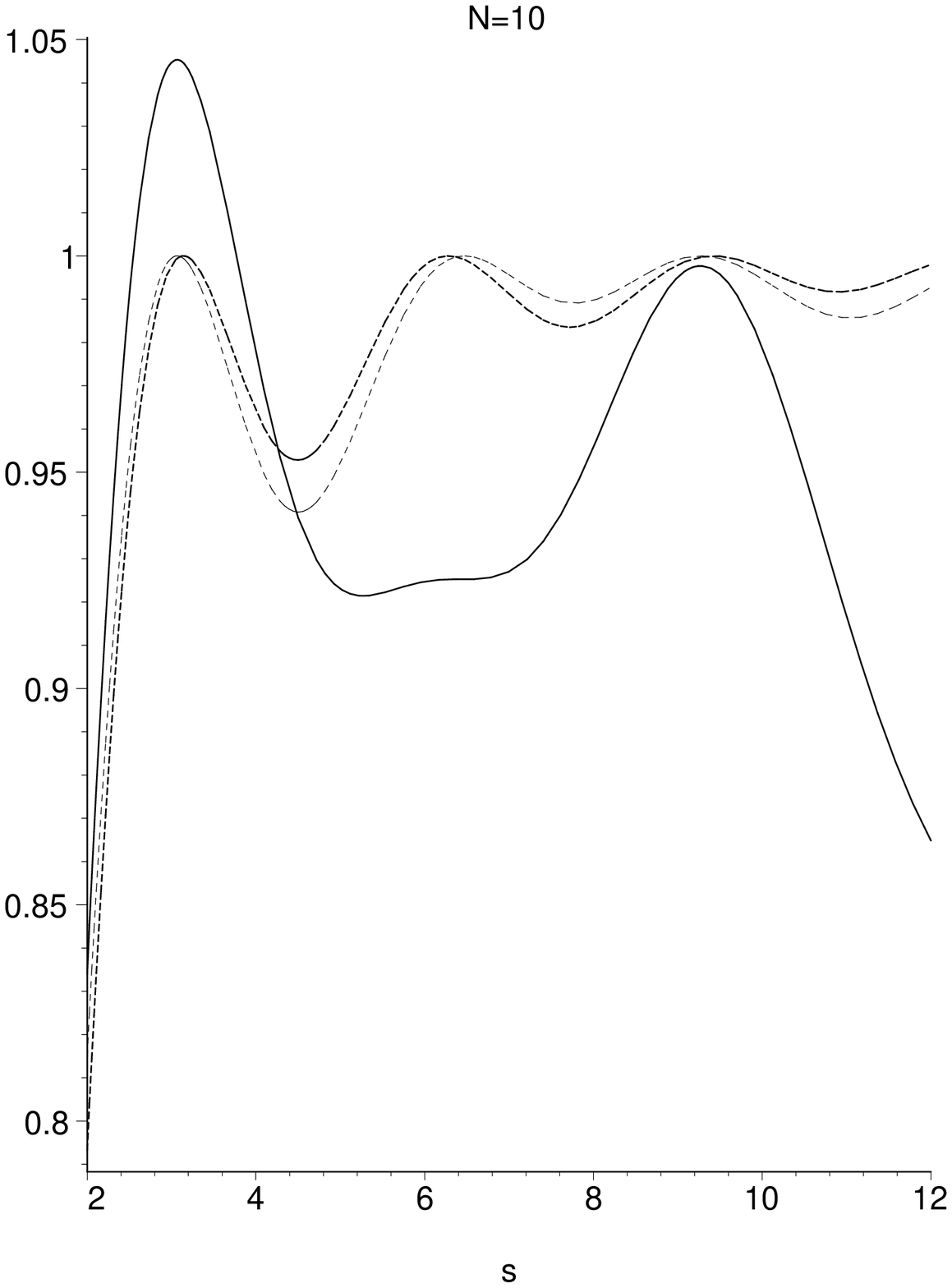}
\vspace{1 cm}
\hspace{3.0 cm}
\epsfxsize=6 cm
\epsfysize=6 cm
\epsfbox{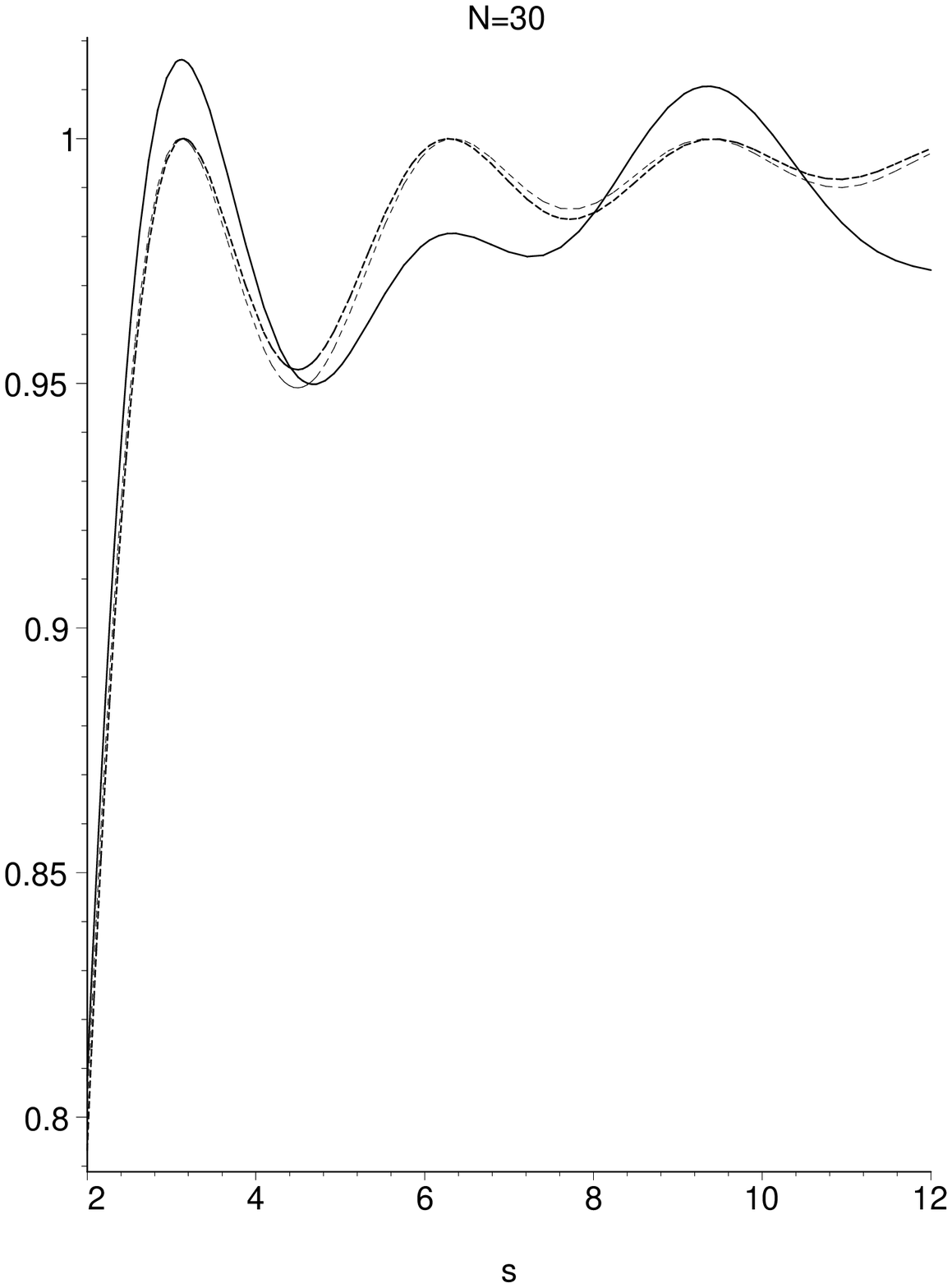}

\vspace{2 cm}
\noindent
\hspace{-2.0 cm}
\epsfxsize=6 cm
\epsfysize=6 cm
\epsfbox{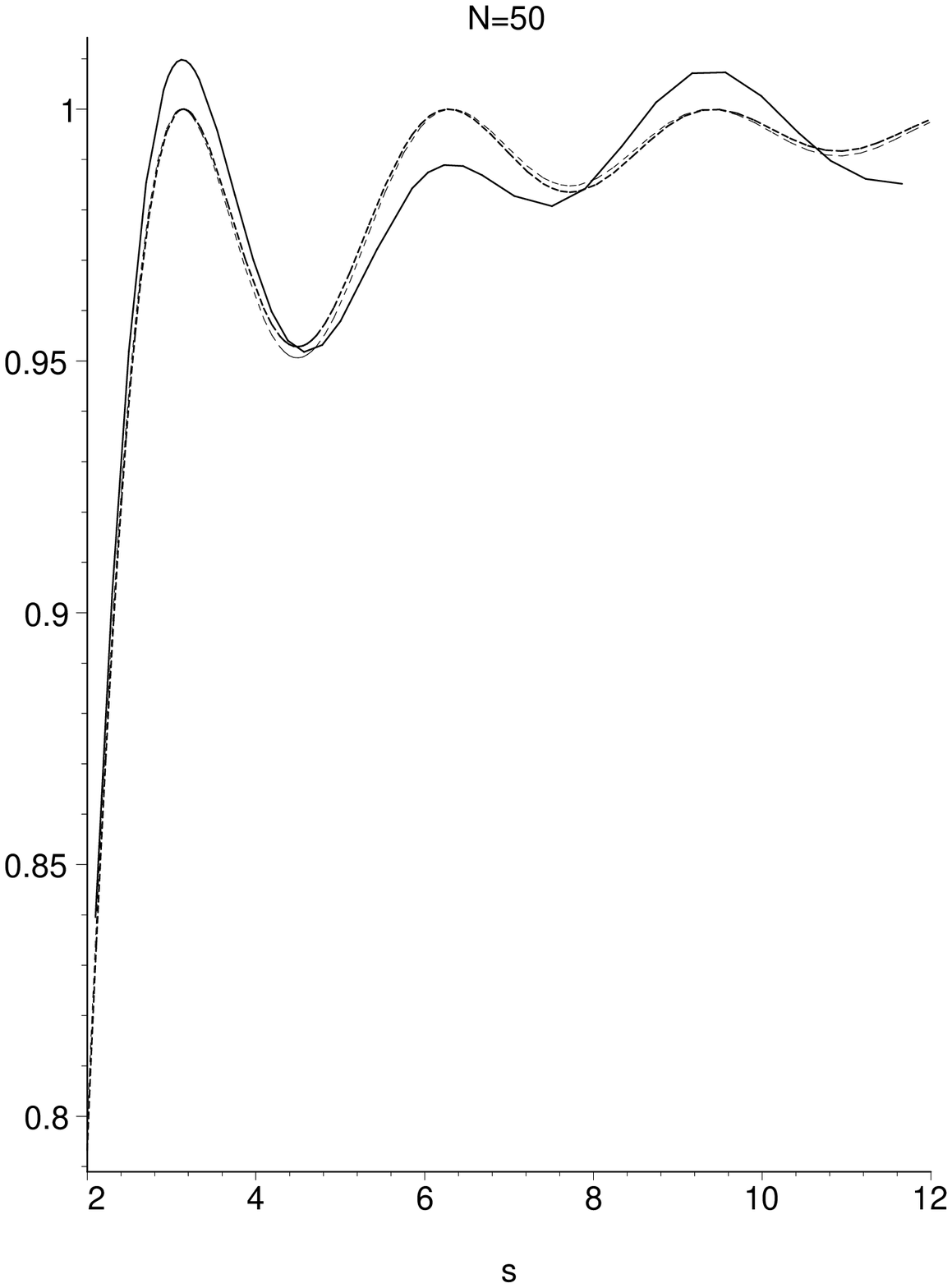}
\vspace{1 cm}
\hspace{3.0 cm}
\epsfxsize=6 cm
\epsfysize=6 cm
\epsfbox{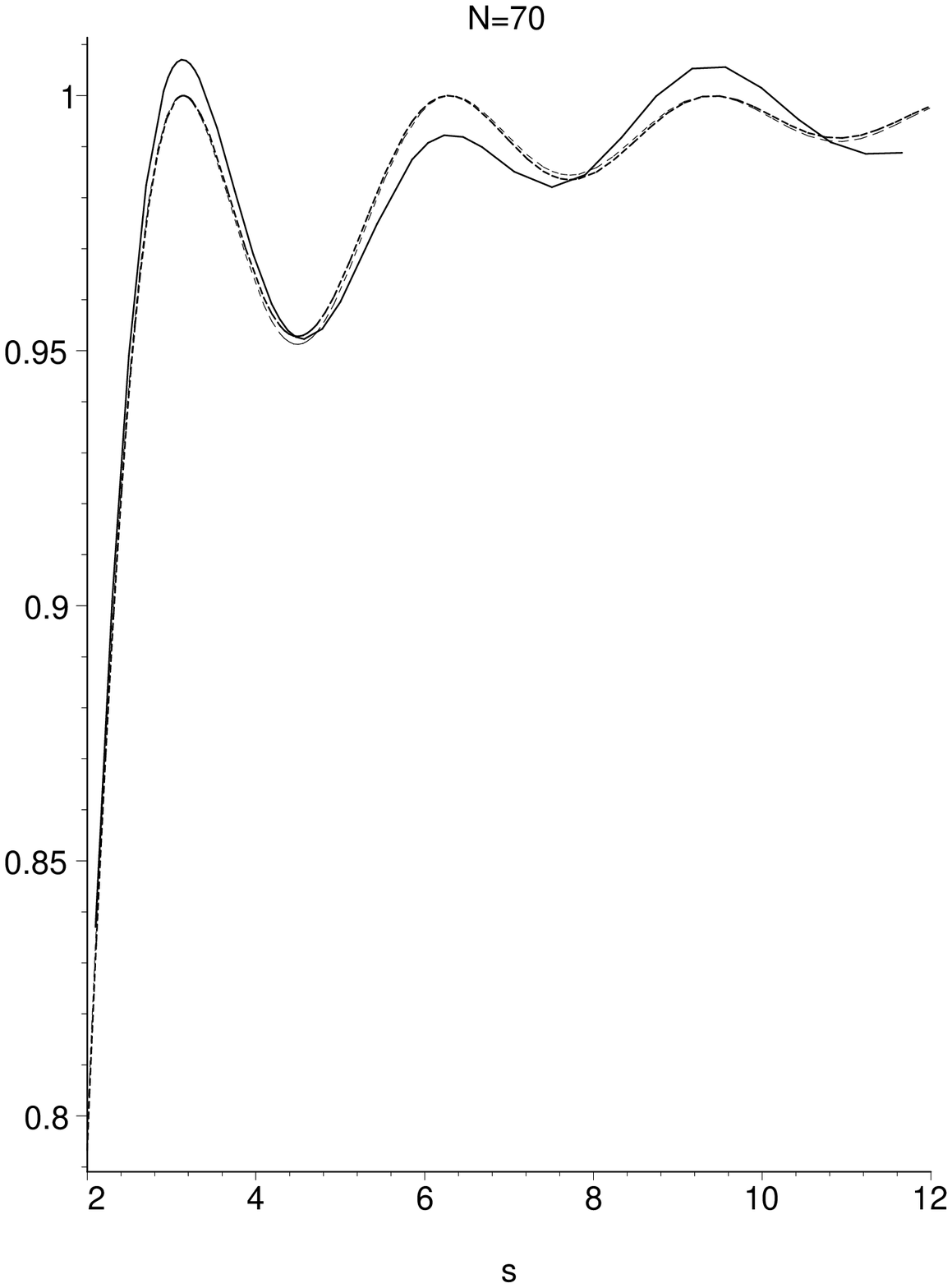}
\caption{In this set of figures  the bold line represents the correlator (\ref{theq}), for $N=10, 30, 50$ and $70$. The dashed line corresponds to the 
$1-(\sin^2 s)/s^2$ law in eq. (\ref{ulaw}), while the dotted line corresponds to the $N$ DOS-DOS correlator obtained in terms of orthogonal polynomials\cite{mehta}. We have considered $\gamma=1$ in the four figures.}
\label{fig2}
\end{figure}

To show this effect in more detail we present 
in figure \ref{fig2} a set of figures in the range $s\in [2,12]$ where  we display a detailed view of the universal law and the convergence of our finite $N$ correlator, by considering $N=10$, $30$, $50$ and $70$.
We can see that for larger values of $N$ more oscillations are accommodated, clearly showing the onset of the asymptotic universal law. We also note in this figure, that in a
region close to the origin the convergence is very fast. The parameter $\gamma$, that characterizes the GUE, has been taken equal to one, however, the same matching has been also verified when $\gamma$ is changed.
In this set of figures, we have also displayed the corresponding finite $N$ DOS-DOS correlators obtained in terms of orthogonal polynomials, as given in Mehta's book \cite{mehta}. Of course, the fitting of these polynomials with the universal law is also improved as $N$ increases. Although
both curves converge to the asymptotic universal law, the finite $N$ results for $(2\gamma N^2)^{-1}\, \Re \, (C(\chi))$ do not coincide with the finite $N$ exact DOS-DOS correlators. This is expected; as we have seen in section \ref{s5}, both quantities differ by terms which only become zero in the $N\to \infty$ limit.

\section{Boson-Fermion Symmetry and Four-Point $O_m$ Correlators in QH Systems
\label{bfs}}

In this section, we will consider a planar system of electrons in the presence of a magnetic 
field $B$ and impurities represented by a random potential $V({\bf x})$. The corresponding 
Hamiltonian operator is,
\begin{equation}
H=H_0 + V
\makebox[.5in]{,}
H_0=\frac{1}{2m}({\bf p}-e{\bf A})^2,
\end{equation}
where $A$ is the vector potential  (we will use the symmetric gauge). 

Here, we can also introduce the quadratic operator $O_m$ according to eq. (\ref{mm1}).
As discussed, because of Wick's theorem, the four-point correlators associated with $O_m$ 
will encode important information about the mixed retarded and advanced products appearing in the 
density of localized states in eq. (\ref{loc}). In a similar way to eq. (\ref{Sact}), 
we can introduce a single superfield with components $\varphi({\bf x})$, $\psi({\bf x})$ 
whose action is given by,
\begin{equation}
S_{m}=\frac{\lambda}{2}\int d^2{\bf x}\,
(\overline{\varphi}\,O_{m}\,\varphi +\overline{\psi}\,O_{m}\,\psi),
\end{equation}
thus leading to a properly normalized representation of the four-point correlators 
containing $G_m$ Green's functions.

For a very strong magnetic field, the transitions between Landau levels are suppressed. 
Then, if we are interested in the case where the first Landau level is filled, we can consider 
the projection,
\begin{equation}
\varphi=e^{-\frac{1}{4}\kappa^2 z\bar{z}} u(z)
\makebox[.5in]{,}
\psi=e^{-\frac{1}{4}\kappa^2 z\bar{z}} v(z),
\end{equation}
where $z=x+iy$ and $\kappa =eB/\hbar$. 

The action $S_m$ for the projected fields can be written as,
\begin{equation}
S_{m}=\frac{\lambda}{2}\int dz d\bar{z}\, 
\left[ \left( V+\frac{\hbar eB}{4m}-E\right)^2-(r/2+i\eta)^2 \right] \alpha(z,\bar{z})
\makebox[.5in]{,}
\alpha(z,\bar{z})=e^{-\frac{1}{2}\kappa^2 z\bar{z}}(\overline{u}u + \overline{v}v).
\label{QHSm}
\end{equation}
We would like to underline that eq. (\ref{QHSm}) is local in the random potential 
(containing up to quadratic terms). Then, if the potentials are uncorrelated at different points, 
the corresponding average will lead to an effective action which is local in the field 
$\alpha(z,\bar{z})$,
\begin{equation}
S_{m}^{eff}=\frac{\lambda}{2}\int dz d\bar{z}\, g(\alpha),
\end{equation}
where $g(\alpha)$ depends on the type of disorder. 

Then, we will be able to follow refs. \cite{ID,Gross}, defining a superfield 
$\Phi(z,\theta)=u(z)+1/\sqrt{2}\, \kappa \theta v(z)$ to write,
\begin{equation}
\alpha(z,\bar{z})=\frac{2\pi}{\kappa^2}\int d\theta d\bar{\theta}\,
e^{-\frac{1}{2}\kappa^2 (z\bar{z}+ \theta \bar{\theta})}\Phi \bar{\Phi},
\end{equation}
(the normalization for the Grassmann measure is $\int d\theta d\bar{\theta}\, e^{-\theta \bar{\theta}}=1/\pi$)  
and construct a supersymmetric representation,
\begin{equation}
S_{m}^{eff}=\frac{\lambda}{2}\frac{2\pi}{\kappa^2}
\int dz d\bar{z}d\theta d\bar{\theta}\, e^{-\frac{1}{2}\kappa^2 (z\bar{z}+ \theta \bar{\theta})}h(\Phi \bar{\Phi})
\makebox[.5in]{,}
h(\alpha)=\int_0^\alpha \frac{d\beta}{\beta} g(\beta),
\end{equation}
with the remarkable boson-fermion symmetry appearing in ref. \cite{Gross}, when computing the total density of states
in QH systems.
This is an interesting result as this is the underlying symmetry behind the exact expression for the total density. 

As discussed in that reference, if the traditional approach to treat mixed retarded and advanced correlators is followed,
the boson-fermion symmetry is lost because of the superfield doubling used to represent each prescription.
Here, we have seen that this important symmetry is restored if the single superfield representation for mixed correlators 
is instead considered.

\section{Conclusions and Perspectives}
\label{s6}

In this work we have introduced a new representation for mixed retarded and advanced 
four-point correlators. These quantities are the relevant ones when studying 
localization and transport properties in disordered systems. 
On the other hand, these are precisely the kind of correlators which are 
more difficult to deal with. This comes about from the mixed prescription
involved, which leads to a doubling of the field copies or the superfield
representation in the traditional approaches.

Here, we have considered a single Green's function
$G_{m}$ for an operator $O_m$, quadratic in the system's Hamiltonian, which
already contains the abovementioned mixing. In this manner, we have been able to encode the nontrivial information in four-point correlators essentially using a single prescription.
In general, this $O_m$ correlator differs from the mixed four-point correlator or the 
density-density correlator by only retarded and only advanced terms, which can be computed by standard procedures.

These ideas have been tested in the simpler context of Random Matrix theory. For
this aim, we have obtained in section \ref{s5}, in the $N\to \infty$ limit,
a direct relationship between the density-density correlator,
\[ 
\langle \hat{\rho}(E_1)\hat{\rho}(E_2)\rangle/(\rho(E_1)\rho(E_2)),
\]
and our four-point correlator for $O_m$ operators, whose GUE average has been computed in section \ref{s3}. Our closed expression (\ref{theq}) for the (finite $N$) $O_m$ correlator clearly shows the onset of the well known asymptotic level spacing correlation.

Then, we have seen that the single superfield method is an interesting alternative to be considered when studying disorder. In particular, in the context of quantum Hall systems, we have shown in section \ref{bfs} that the boson-fermion symmetry is restored for correlators encoding information about mixed retarded and advanced Green's functions, thus opening
the possibility of improving our understanding of localization \cite{inp}.

\acknowledgments
The Conselho Nacional de Desenvolvimento Cient\'{\i}fico e
Tecnol\'{o}gico (CNPq-Brazil), the Funda{\c {c}}{\~{a}}o de Amparo
{\`{a}} Pesquisa do Estado do Rio de Janeiro (FAPERJ), CAPES and SR2-UERJ
are acknowledged for the financial support. 
D.\ G.\ Barci would like to acknowledge the ``Abdus Salam international center for theoretical physics, ICTP'', for kindly hospitality during part of this work.

\section*{Appendix}

Here we show how to compute traces of matrices which are functions of
$K=\varphi \otimes \overline{\varphi} -\psi \otimes 
\overline{\psi}$. For a function $f(x)$, with a corresponding expansion 
$f(x)=\sum_n f_n x^n$, we have, $tr f(K)=\sum_n f_n \,tr K^n$.
On the other hand, a direct computation shows that,
\begin{eqnarray}
\,trK&=&\alpha-\beta \nonumber \\
\,trK^2&=&\alpha^2+2\mu \overline{\mu}-\beta^2\nonumber \\
\,trK^3&=&\alpha^3+3(\alpha+\beta)\mu \overline{\mu}-\beta^3,
\end{eqnarray}
where,
\begin{equation}
\alpha=\varphi \cdot \overline{\varphi}~~~,~~~
\beta=\psi \cdot \overline{\psi}~~~,~~~
\mu=\varphi \cdot \overline{\psi}.
\label{inva}
\end{equation}
It is easy to see the general form of the trace for a given power of K,
\[
\,trK^n=\alpha^n+n(\alpha^{n-2}+\alpha^{n-3}\beta+\alpha^{n-4}\beta^2
+\dots+\alpha^2\beta^{n-4}+\alpha\beta^{n-3}+\beta^{n-2})\mu
\overline{\mu}-\beta^n, 
\]
or in an equivalent form,
\begin{equation}
tr K^n=\alpha^n+n\frac{\alpha^{n-1}-\beta^{n-1}}{\alpha-\beta}\mu
\overline{\mu}-\beta^n\makebox[.5in]{,} n\neq 0.
\end{equation}
Then, we arrive to a formula that gives the trace of a function of $K$ just in
terms of the $U(N)$-invariants (\ref{inva}):
\begin{eqnarray}
\,tr f(K)&=&\,tr \left( f_0 I+\sum_{n=1}^{\infty} f_n K^n\right)
\nonumber \\
&=&Nf_0+\sum_{n=1}^{\infty} f_n
\left(\alpha^n+n\frac{\alpha^{n-1}-\beta^{n-1}}{\alpha-\beta}\mu 
\overline{\mu}-\beta^n \right)\nonumber \\
&=&Nf_0+f(\alpha)-f(\beta)+\frac{f'(\alpha)-f'(\beta)}{\alpha-\beta}
\mu \overline{\mu}.
\label{tdeFK}
\end{eqnarray}


\end{document}